\documentclass[conference]{IEEEtran}
\usepackage{tikz}
\ifCLASSOPTIONcompsoc
  \usepackage[nocompress]{cite}
\else
  \usepackage{cite}
\fi
\usepackage{graphicx}
\usepackage{listings}
\usepackage{bbm}
\usepackage{amsmath}
\usepackage{url}
\usepackage{array}

\newcommand*\rot{\rotatebox{90}}

\begin{document}

\def\cm{\tikz\fill[scale=0.4](0,.35) -- (.25,0) -- (1,.7) -- (.25,.15) -- cycle;} 

\title{Julia Implementation of the Dynamic Distributed Dimensional Data Model}

\author{\IEEEauthorblockN{Alexander Chen, Alan Edelman}
\IEEEauthorblockA{
Massachusetts Institute of Technology\\
Cambridge, MA, U.S.A.\\}
\and
\IEEEauthorblockN{Jeremy Kepner, Vijay Gadepally}
\IEEEauthorblockA{MIT Lincoln Laboratory\\ 
Lexington, MA, U.S.A.\\}
\and
\IEEEauthorblockN{Dylan Hutchison}
\IEEEauthorblockA{University of Washington\\ 
Seattle, WA, U.S.A.\\}}

\maketitle

\begin{abstract}
  Julia is a new language for writing data analysis programs
that are easy to implement and run at high performance.
  Similarly, the Dynamic Distributed Dimensional Data Model (D4M) aims to clarify data analysis operations while retaining strong performance.  D4M accomplishes these goals through a composable, unified data model on associative arrays.
  In this work, we present an implementation of D4M in Julia
and describe how it enables and facilitates data analysis.
  Several experiments showcase scalable performance in our new Julia version
as compared to the original Matlab implementation.
\end{abstract}

\IEEEpeerreviewmaketitle

\section{Introduction}  

Data analysis is the science of converting raw data into insight. 
As the boom of the information age continues, data analysis larger and larger data sets, hence the increased focus on Big Data.  
One of the 4V challenges in Big Data is velocity ~\cite{v4s}.  
In most cases, velocity in big data describes the need for
computational performance to keep up with the rate of data that is generated.  
Developing data analysis programs that produce useful results while
performing efficient analysis is a time-consuming process.  
The development difficulty and cost of creating data analysis
software has been one of the top barriers for companies to develop
analytical solutions ~\cite{russom2011big} and a large reason behind
very high-level languages such as MATLAB~\cite{samsi2010matlab} or Julia.
Further, computational performance is not necessarily guaranteed with
the correctness of the implemented algorithm.


The D4M (Dynamic Distributed Dimensional Data Model) technology
provides a composable toolbox to operate on structured or unstructured data~\cite{kepner2013d4m}.  
The MIT SuperCloud Team has been using D4M to prototype analysis solutions under their Common Big Data Architecture (CBDA) ~\cite{gadepally2014big}.  
D4M leverages concepts in linear algebra and signal processing to enable a mathematically rich representation of associative arrays~\cite{kepner2014adjacency}.  
A practical application tested with D4M is a comparison between two sets of biomedical data, more specifically ribonucleic acid (RNA) sequence snippets.  By transpose matrix multiply by an matrix of known sequence ID and a matrix of unknown sequence ID, D4M is able to display the correlation between the two sets of sequences ~\cite{dodson2014genetic}.  
D4M represents data in Associative Arrays (Assoc) which is shown in
Figures 1 and 2. D4M's Associative Array (Assoc) differs from a
DataFrames objects as Assoc utilizes sparse matrix assumptions.  
Sparse matrix representation easily converts to the graph representation of edges and vertices ~\cite{konig1936theorie, konig1931graphen}.
Also, D4M provides linear algebraic and associative mathematics operations unlike those available for Dataframe.jl ~\cite{dataframejl}, which provide mainly tools with statistical analysis in mind.

\begin{figure}[h!]
\center
\begin{tabular}{c|cccc}
   & Alice  & Bob &  Casey & Joe \\
  \hline
  0730 & 30 &  & 30 &  \\
  1145 &  & 60 &  &  60 \\
  1400 &  & 15 & 15 &  \\ 
\end{tabular} 
\caption{General Tabular Data Representation}
\end{figure}
 
\begin{figure}[h!]
\center
\shortstack{
  \begin{tabular}{c|ccc}
  r & 0730 & 1145 & 1400 \\
  \hline
  i & 1 & 2 & 3  \\
  \end{tabular} \\
  \begin{tabular}{c|cccc}
  c & Alice & Bob & Casey & Joe \\
  \hline
  j & 1 & 2 & 3 & 4 \\
  \end{tabular} \\
  \begin{tabular}{c|ccc}
  v & 15 & 30 & 60 \\
  \hline
  k & 1 & 2 & 3 \\
  \end{tabular} \\
  $\begin{bmatrix}
  2 & & 2 & \\
  & 3 & & 3 \\
  & 1 & 1 & \\
  \end{bmatrix}$
}
\caption{Associative Array Representation}
\end{figure}

Julia is a new language developed to make high performance, high-level dynamic programs ~\cite{bezanson2014julia}.  
To do so, the developers of Julia focused on the two-language problem: the excess of having both a high-level language with ease of development but under performance and a low-level language with performance but difficulty in development.
In Julia, there are both state of the art numeric computation libraries and a state of the art Just-in-Time (JIT) compiler built on Low Level Virtual Machine (LLVM) ~\cite{DBL:journals/corr/abs-1209-5145}.
The principal idea is that/ Julia's low-level functionality needs to be optimized.  This principle is so that the user will not resort to use other languages as low-level building blocks.  
By leveraging the selected chain of modern programming language technologies within Julia, the Julia community has been rapidly expanding the high-level functions of Julia without compromising in performance. 
Julia has proven to be effective in high performance computing, although some of its user interface elements, such as the debugger, still have room to grow  ~\cite{eadline2016hpcjulia,nowozin2015scientificjulia}.
As the Julia ecosystem has rapidly matured, advanced and efficient libraries have been released in Julia ~\cite{juliapulse}.  
For example, the Julia for Mathematical Programming (JuMP) library provides researchers high-level tools to implement optimization problems while delivering performance faster than similar open source tools, and comparable to commercial options ~\cite{LubinDunningIJOC}.  
These expansive libraries allow Julia to cover areas traditionally
uncoverable by a single library built on a single language, due to
language specialization and optimization. Thus, Julia's developers could find performance enhancements with this new scope.

Building D4M on top of Julia enables the already flexible D4M analysis tool to be more effective and optimized without sacrificing prototyping convenience.  
Likewise, building D4M for Julia enables other unique forms of data analysis approaches enabled on Julia.  
Some linear algebraic data analysis approaches have been implemented on Julia, and shown good results in achieving both performance and ease ~\cite{shah2013novel}.

\section{Approach}

In the original D4M-Matlab implementation, D4M was developed to be integrated within the Matlab environment~\cite{MATLAB:2010}.  
Because associative array is represented as a matrix, a lot of native matrix operations in Matlab are well suited to represent equivalent operations for associative arrays.  
Similarly, Julia as a numerical computation language, has a wealth of operators and operations ready for matrix.  
This implementation of D4M in Julia is called D4M.jl.  
During the implementation of D4M.jl we followed Julia syntax where possible.  
For example, for indexing elements within an associative array, instead of utilizing the normal bracket as in Matlab, the square bracket is used as per Julia conventions.  
Although this would cause trouble while converting from D4M-Matlab scripts into D4M.jl, this design approach focuses on clarity when developing within Julia.  
As a result, when implementing new algorithms and programs using D4M.jl in combination with other Julia libraries, the developer would not need to hop between Matlab syntax for D4M and Julia syntax.

\subsection{Strict Typed Methods}  
One of Julia's key features is a well-defined type system.  
As written in the Julia paper, ``The more the computer knows about this data, the better it is at executing operations on that data'' \cite{bezanson2014julia}.  

Our Julia implementation leverages well-defined types.  
For example,  associative array indexing  
(called \texttt{getindex} in Julia and \texttt{subref} in Matlab) relies on a type system to build code clarity.  
With Julia, D4M has the option to overload the \texttt{getindex} with strongly typed inputs.  
At the base level, \texttt{getindex} receives an array of indices to output in each dimension.   
Subsequent type inputs then build on top of this base level.   
Figure 3 illustrates how regex indexing is added to the indexing methods.  
Instead of detecting the types of input within the method by giving a \texttt{getindex} without static type checking, new input types are statically typed in new \texttt{getindex} methods that are built on top of a collection of implemented \texttt{getindex}.
This allows the added regex format to resolve into a base level integer index \texttt{getindex}, while declaring how the new regex input is reduced to a more fundamental input type.  

\begin{figure}[h!]
\begin{lstlisting}[frame=single]
toPrevTypeRow(i::Regex,A::Assoc) =
  find( x -> ismatch(i,x),A.row)
toPrevTypeCol(j::Regex,A::Assoc) =
  find( x -> ismatch(j,x),A.col)
	
getindex(A::Assoc, i::Regex, j::PreviousTypes) = 
  getindex(A, toPrevTypeRow(i,A), j)
getindex(A::Assoc, i::PreviousTypes, j::Regex) = 
  getindex(A, i, toPrevTypeCol(j,A))
getindex(A::Assoc, i::Regex, j::Regex) = 
  getindex(A, toPrevTypeRow(i,A), toPrevTypeCol(j,A))
\end{lstlisting}
\label{fRegexIndex}
\caption{Regex \texttt{getindex} reduced to index list \texttt{getindex}}
\end{figure}

This style enables clear declaration for each new input type to be either reduced to a basic array of indices or handled separately as a unique case.
Moreover, it allows for a more concise code, halving the code size for the function while providing the same operation.  
Within the D4M-Matlab, the syntax for \texttt{getindex} with row/column values starts from a dynamically typed object and classifies them through conditionals within the method.  
In contrast, D4M.jl strictly declares this as an array of values.  
However, a contradiction is quick to surface.  
How would D4M.jl resolve a conflict between an array of integer indices and an array of integer row/column values?  
If the user attempts to getindex of the first row or the row numbered 1 (there could be a row numbered -1), there would be collisions in representation.  
The issue would be that the getindex declarations would both be \texttt{Array\{Integer\} }, and the given would be exactly the same while the expected output is completely different.  
By utilizing the precise type declarations that Julia allows, D4M.jl declares all arrays of row/column values as \texttt{Array\{Union\{ AbstractString,Number\}\}}.  
This union type of string and number gives a precise type for all D4M.jl to pass an array of row/column or val mapping between operations.  
Thus, in our example, users that executed a filter on the list of row values and then tried to pass to Assoc to select the rows, wouldn't encounter this problem.  
When developing analysis software around D4M, this also allows users to build more precise type declaration for what D4M.jl expects.  

\subsection{DataFrames and Assoc}
Within Julia, there is already a current representation of 2D tabular data: DataFrame.jl ~\cite{dataframejl}.  
D4M and Dataframe differ primarily in D4M's linear algebraic interface, which contains a wealth of composable computations.  
The figure below compares D4M with the common Dataframe implementations: pandas ~\cite{mckinney2012pandas} and R's DataFrame ~\cite{R}.  
Most of the DataFrame implementations are intentionally quite similar, because DataFrame serves as a good representation of statistical data.  
However, D4M's approach from a more graph-oriented thinking differs it from DataFrame.  
D4M's main purpose is to provide clean, composable operations.  
The associative array is a data representation to allow high performance with those operations.  
Because the approach is graph-oriented, there are more chances of the data becoming sparse as compare with a pure statistical approach.  The implicit sparse storage with associative array aids in the performance of D4M's operations.

\begin{figure}[h!]
\center
\begin{tabular}{c|c|c|c|c}
   x & Alice  & Bob &  Casey & Joe \\
  \hline
  0730 & 30 & 0 & 30 & 0 \\
  1145 & 0 & 60 & 0 &  60 \\
  1400 & 0 & 15 & 15 & 0 \\ 
\end{tabular} 
\caption{DataFrame Representation}
\end{figure}

\begin{figure}[h!]
\center
\begin{tabular}{l|c|c|c|c|}
 & \rot{R.DataFrame} & \rot{Python.pandas} & \rot{DataFrame.jl} & \rot{D4M.jl}\\
\hline
Sparse Storage  & \cm & \cm  & \cm & \cm\\ 
\hline
Composable Compute  & \cm & \cm  & \cm & \cm\\ 
\hline
Composable Query  & \cm & \cm  & \cm & \cm\\ 
\hline
Linear Algebraic Interface &  &   &  & \cm\\ 
\hline
Implicit Sparse Storage  &   &   &  & \cm\\ 
\hline
\end{tabular}
\caption{Technology Comparison}
\end{figure}

\begin{figure}[h!]
\begin{lstlisting}[frame=single]
DataFrame df = DataFrame(A::Assoc)
Assoc A = Assoc(df::DataFrame)
\end{lstlisting}
\caption{Assoc Interfacing with DataFrame}
\end{figure}

In order to allow swift interface between DataFrame and D4M, there are a few assumption issues that need to be resolved.  
This problem can be seen through how both relate to a common format, Comma-Separated Values (CSV).
For Assoc, it is assumed that the first column is always the row names.  
This assumption is due to the explicit requirement that there must be a name for each row, just like columns.  
On the other hand, DataFrame treats all data points within the columns equally.  
It requires that all columns have a name, and thus it automatically assigns columns without names a temporary variable $x$.  
During conversion, this difference in assumption has to be taken into account.  
How should rows be named when converting from Dataframe to Assoc, or how should the row names be stored converting from Assoc to Dataframe.
This is also the reason why utilizing DataFrame to execute D4M operations would be inefficient.

\subsection{Utilizing Libraries Developed by the Julia Community}
Compared to Matlab, Julia core is missing some of the functions that a Matlab user would normally expect.  
An example of this is plotting and save/load current data.  
Thankfully, the ever growing Julia community has implemented many of these essential functions in Julia.  
For development example scripts of D4M.jl, libraries that, allows save and load data structure directly to hard drive ~\cite{JLDjl} and basic plotting capabilities ~\cite{PyPlot}, enables the visualization and other features comparable to the D4M-Matlab counter part.

\subsection{Sorted Set Operation}
In D4M, the totally ordered set of string and number to index mappings are stored as sorted arrays with unique elements ~\cite{birkhoff1940lattice}.  
To achieve total orderness, the data structure must hold the comparability property.  
Thus, a comparable has been added for Julia's AbstractString and Numbers when D4M is enabled.  
This will result in D4M always sorting AbstractString to be less than Numbers, which is feasible only because AbstractString and Numbers stand in two completely different branches of types.  
Within Julia's sophisticated type system, the only possibility that a type can be both AbstractString and Number is by type union.  
Within the union, orderness is resolved by the underlying data type of the element.  

Because of the data representation of the mappings as ordered array, an efficient set of sorted set operation primitives greatly aid in the performance of D4M operations.  
These include union, intersects, and mapping.  
Similar to D4M-Matlab, D4M.jl stores its index mapping in the sparse matrix as a sorted array.  
The operation primitives utilize the orderness and uniqueness of the array to remove redundant checks and operations.

In sorted intersect and sorted union, two target indices are created.  
As the operations traverse the two arrays, the indices are used to keep track of progression, and the algorithm does not need to look back, allowing $O(n)$ performance instead of $O(n^2)$ in a naive approach.  
An iterator for the arrays will have the same effect as a using target index.  
The development runtime comparison figures below (figure 7 and 8) show the effect that the intersect and union had on a naive approach.

With sorted intersect and sorted union, these bring the D4M operation performance to the level of Matlab.  
To speed up the D4M operation further, a new primitive has been isolated and improved: sorted mapping.  
The purpose of mapping operation is to map the first given set to the indices of the matching elements in a second given set.  
In D4M-Matlab, this is done by StrSubsRef, which is used as a String sub-referencing.  
The Matlab operation includes the ability to sub-reference strings with Regex, and thus relaxed the requirement of the second argument the array.  
In D4M.jl, the orderliness of the input array is used, and thus a slight performance boost is gained, similar to sorted union and intercept.  
The figures below show the performance boost gained during development by leveraging these set operations.

\begin{figure}[h!]
\includegraphics[width=0.5\textwidth] {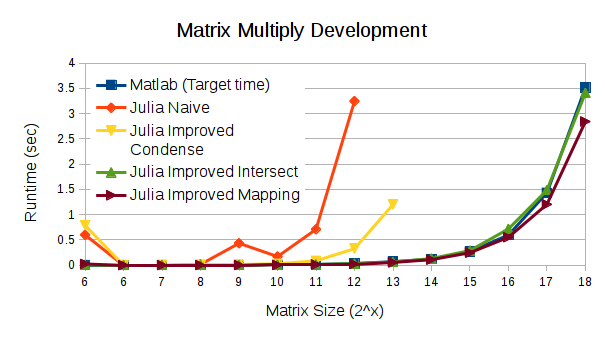}
\caption{Matix Multiply Runtime with Incremental Versions}
\end{figure}

\begin{figure}[h!]
\includegraphics[width=0.5\textwidth] {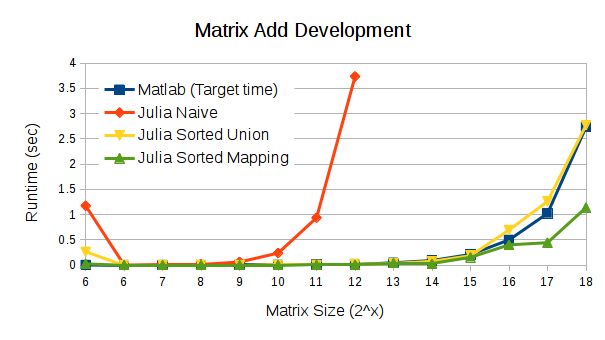}
\caption{Matix Add Runtime with Incremental Versions}
\end{figure}

\section{Performance}

\subsection{Setup}
The performance is to be measured by the matrix performance scripts within the D4M-Matlab example code.  
We have chosen four main methods that have been optimized in the original D4M versions: Matrix Multiply, Concatenated Key Matrix Multiply, Concatenated Value Matrix Multiply, and Matrix Addition.

Matrix Multiply tests the spare matrix indexing and the native sparse matrix multiply in Julia.  
In this case, the test maps the A and B input associative array to an intersect set of the two, and the matrix multiplies the two's sparse matrix.  
Note that this is a purely numerical operation.  
Thus, the string valued associative array is converted to logical, where all nonempty values are 1.  
The equation below describes the operation.

\begin{equation*}
C_{i,j} = \sum^K_k ( A_{i,k} * B_{k,j} )
\end{equation*}

Concatenation Key Matrix Multiply(CatKeyMul) tests the ability of the associative array to do matrix multiply.  
Unlike the matrix multiply in the previous paragraph, in this case, the definition of multiply is being redefined.  
This requires the function to re-implement matrix multiply at a higher level.  
In essence, this function results in values that are a list of indices of A's Column and B's Row that multiplication was successful on.  
Equation below describes the operation.

\begin{equation*}
C_{i,j} = Concat( k ) ; \forall A_{i,k}B_{k,j} \neq 0 ,  k \in 1,\ldots, n
\end{equation*}

Concatenation Value Matrix Multiply (CatValMul) is similar to Concatenation Key Matrix Multiply.  
In this new case, we are storing the values in both A and B instead of just the index which the the multiply coincide at.  
This is a slightly more rigorious version of the CatKeyMul, and similarly the matrix multiplication needs to be implemented.
The equation below describes the operation.

\begin{equation*}
C_{i,j} = Concat( A_{i,k},B_{k,j} ) ; \forall A_{i,k}B_{k,j} \neq 0 ,  k \in 1,\ldots, n
\end{equation*}

Matrix Addition tests element-wise operations on associative array.  
There are two main element-wise operations matrix addition performs based on the type of associative array.  
If both associative arrays are string valued, matrix addition will return a union of the two.  
If not, the method will return the addition of the numeric values.  
If one of the associative array is string-valued, they will be converted to logical.  
The equation below describes the operation.

\begin{equation*}
C_{i,j} =  
\begin{cases}
max(A_{i,j},B_{i,j}), & \begin{tabular}{@{}l@{}}\text{\emph{if} isa(Val(A),Array(String))} \\ \text{\&\& isa(Val(B),Array(String))} \end{tabular} \\
A_{i,j}+\mathbbm{1}_{B_{i,j} \neq 0}, & \begin{tabular}{@{}l@{}}\text{\emph{if} isa(Val(A),Array(String))} \\ \text{\&\& ! isa(Val(B),Array{String}))} \end{tabular} \\
B_{i,j}+\mathbbm{1}_{A_{i,j}\neq 0}, & \begin{tabular}{@{}l@{}}\text{\emph{if} ! isa(Val(A),Array(String))}\\ \text{\&\& isa(Val(B),Array(String))} \end{tabular} \\
A_{i,j}+B_{i,j}, & \emph{otherwise}
\end{cases}
\end{equation*}

\subsection{Measurement}

We conducted five runs of each operation on two separate platforms.
The first data point of each run, at data size $2^6$,
shows the performance of Julia's just-in-time compiler
since Julia compiles code on the fly the first time it runs an operation.
We expected that just-in-time compiling would hamper performance for small matrix sizes,
but it was an insignificant overhead on large matrix sizes.
For subsequent data points, we take the median of the five runs as the result of our experiment.  

We ran our tests on two platforms. 
The first platform is a laptop with specifications of a CPU of i5-3317U Processor (1.7/2.6 GHz 1600MHz 3MB) and a ram setup of 1x8GB PC3-12800 DDR3L SDRAM 1600 MHz.  
The second platform is one node of the MIT Lincoln Laboratory TXE1 Supercomputing cluster.  
In all instances, D4M is run in single thread to test the raw operation performance. 
Measurements are taken in float point operations per second (Flops).  
The measure is done by the calculation of the expected reward float point operation divided by the measure run time.  
Thus, faster operations have a higher measured Flops.

\begin{figure}[h!]
\includegraphics[width=0.47\textwidth] {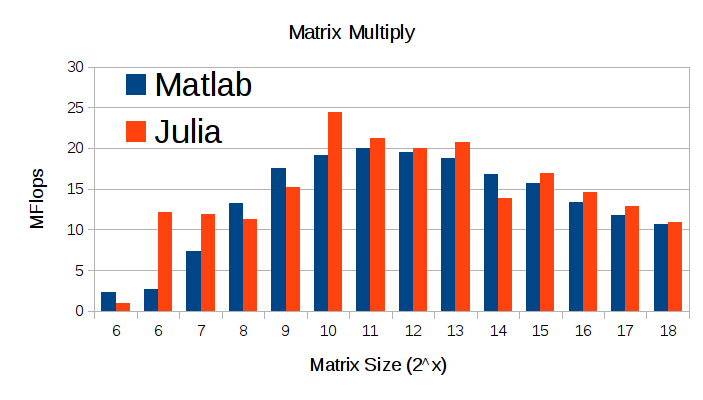}
\caption{Matrix Multiply Mflops on Laptop\label{fMMLaptop}}
\end{figure}

\begin{figure}[h!]
\includegraphics[width=0.47\textwidth] {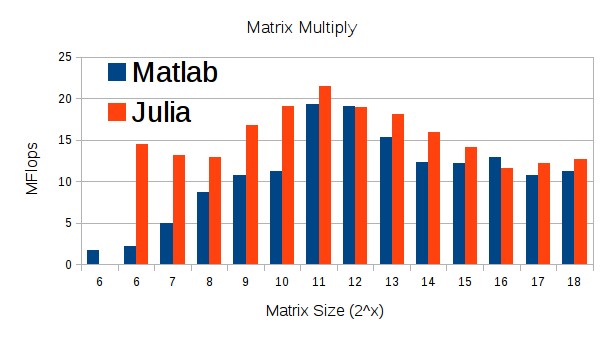}
\caption{Matrix Multiply Mflops on TXE1 node\label{fMMTXE1}}
\end{figure}

The first operation is Matrix Multiply.  
Matrix Multiply heavily relies on the sparse matrix multiply native to Julia.  
Other than the sparse matrix multiply, the D4M Matrix Multiply uses the set operations to correctly match the row and column values of the two input associative arrays, which is the reason why the aforementioned set operations are critical in D4M performance.  
In the Laptop measurement, the runs with smaller matrix size have some trouble, but this is most likely due to the Laptop's platform constant overheads than to algorithmic bugs.  
In contrast, on TXE1, D4M.jl performs consistently better than D4M-Matlab.  
Figure \ref{fMMLaptop} and \ref{fMMTXE1} display the recorded measurements.

The second operation is CatKeyMul.  
Even though operation-wise quite similar to Matrix Multiply, the multiplication process of the CatKeyMul have to be coded explicitly instead of relying on multiplication process available in Julia core.  
To override underlying matrix multiplication process in Julia would create too many conflicts for just this operation.  
Though various other approaches have been attempted in Julia, for example using BigNum to represent the bitvector of the set, the approach coded in Matlab-D4M is indeed an optimal method to conduct Sparse Matrix Multiplication with element concatenation.  

\begin{figure}[h!]
\includegraphics[width=0.47\textwidth] {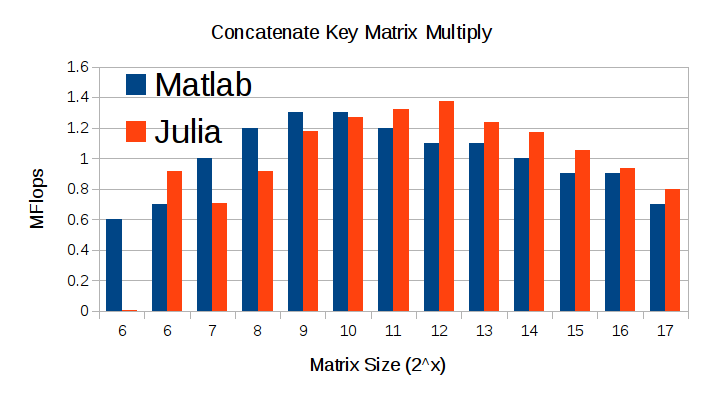}
\caption{Concatenation Key Matrix Multiply Mflops on Laptop\label{fCKMLaptop}}
\end{figure}

\begin{figure}[h!]
\includegraphics[width=0.47\textwidth] {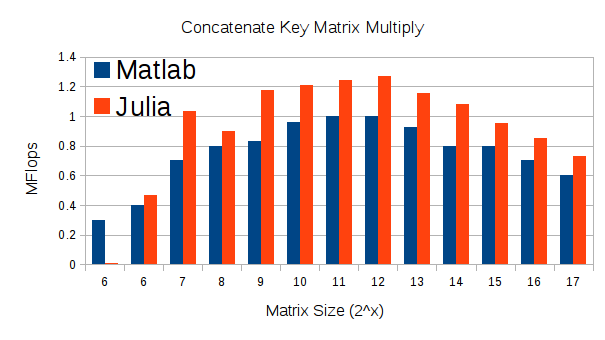}
\caption{Concatenation Key Matrix Multiply Mflops on TXE1 node \label{fCKMTXE1}}
\end{figure}

In the end, though the approaches are very similar to Matlab-D4M, 
the Julia version outperforms the Matlab version 
due to Julia's strong sparse matrix support.
However, as in the case of Matrix Multiply, 
the Matlab version does run faster for small matrix sizes.  
Figures \ref{fCKMLaptop} and \ref{fCKMTXE1} display the recorded measurements.

\begin{figure}[h!]
\includegraphics[width=0.47\textwidth] {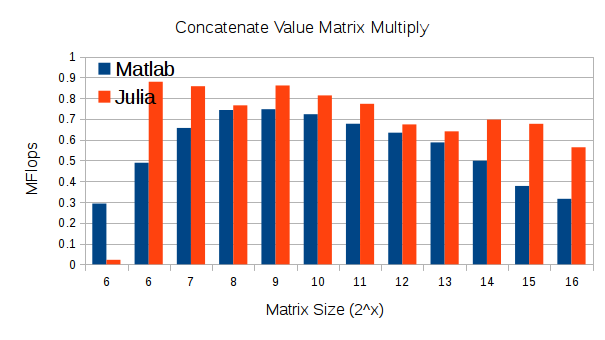}
\caption{Concatenation Value Matrix Multiply Mflops on Laptop\label{fCVMLaptop}}
\end{figure}

\begin{figure}[h!]
\includegraphics[width=0.47\textwidth] {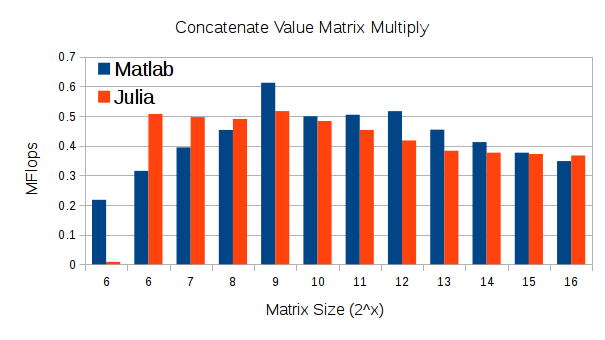}
\caption{Concatenation Value Matrix Multiply Mflops on TXE1 node\label{fCVMTXE1}}
\end{figure}

The third operation is CatValMul.  
CatValMul is very similar to CatKeyMul.  
However, the operation contains more string operations.  
Interestingly, on TXE1, the result seems that D4M.jl is performing worse than Matlab-D4M.  
On comparison with on laptop performance, D4M.jl performance profile is very alike.  
Instead, the Matlab-D4M profile shows the Matlab-D4M is performing better on TXE1.  
This is most likely because of better cache performances, on the TXE1 compare with on the laptop.  
In large matrix size, D4M.jl and Matlab-D4M performs quite close in TXE1.  
The factor speed up the Julia gains could potentially recover when the matrix size is scaled higher.  
Figures \ref{fCVMLaptop} and \ref{fCVMTXE1} display the recorded measurements.

\begin{figure}[h!]
\includegraphics[width=0.47\textwidth] {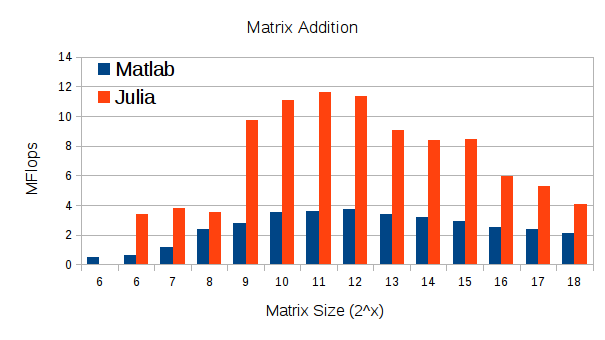}
\caption{Matrix Addition Mflops on Laptop\label{fMALaptop}}
\end{figure}

\begin{figure}[h!]
\includegraphics[width=0.47\textwidth] {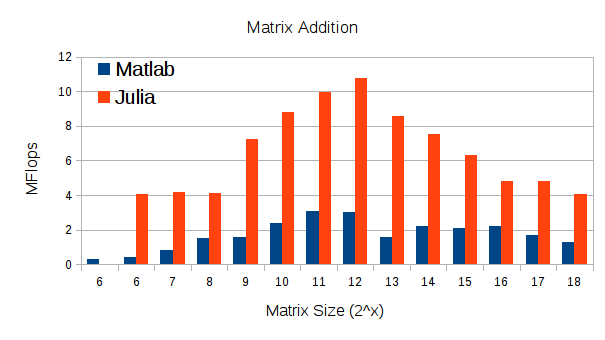}
\caption{Matrix Addition Mflops on TXE1 node\label{fMATXE1}}
\end{figure}

The last operation is Matrix Addition.  
Compare with Matrix Multiply, Matrix Addition relies even more on sorted set operations.  
The ability to quickly find union sets of the two input associative array and also seek the column and row mapping between the two input associative array and the resultant union associative array.  This allows D4M.jl to out perform its Matlab counter part.  
Matrix Multiply seems to have a higher MFlops compare with Matrix Addition.  
This is because the majority of Matrix Addition's operation is on computing the mapping of the associative arrays, instead of just the sparse addition operations.  
Sparse addition operations is the only factor in calculating the reward float point operations, not computing the indice mappings.  
This is true both for D4M.jl and D4M-Matlab.  
Figures \ref{fMALaptop} and \ref{fMATXE1}  display the recorded measurements. 

\section{Summary}

This paper presents D4M.jl, a version of D4M built on top of Julia.  
In the development of D4M.jl, there are factors that brings Julia ahead of Matlab.  
Julia's well structured syntax and data structure allowed stronger performing fundamental operations to be created.  
Thus, this enabled higher level operations such as Matrix Addition for D4M.jl, to be faster than, if not comparable, its optimized Matlab counterpart.  
Because of Julia's expanded capability and libraries, future expansions of D4M's ability to connect to new databases~\cite{gadepally2015d4m} would be an interesting direction.  
D4M.jl is available on a public repo: github.com/achen12/D4M.jl.

\ifCLASSOPTIONcompsoc
  
\else
\fi

\bibliography{mybib}{}
\bibliographystyle{plain}

\end{document}